\documentclass[%
 reprint,
amsmath,amssymb,
aps,
pra,
floatfix,
]{revtex4-2}

\usepackage[dvipsnames]{xcolor}
\definecolor{r}{rgb}{1.,0.,0.}
\definecolor{g}{rgb}{0.,1.,0.}
\usepackage{graphicx}
\usepackage{dcolumn}
\usepackage{bm}
\usepackage{subfigure}
\renewcommand{\selectlanguage}[1]{}

\begin{document}


\title{Microstructure of Polydisperse Colloidal Gels}

\author{Benjamin F. Lonial}
\email{Current address:  Department of Physics, University of California at Santa Barbara,  blonial@ucsb.edu}

\author{Eric R. Weeks}%
 \email{erweeks@emory.edu}
 
\affiliation{Department of Physics, Emory University, Atlanta, GA 30322, USA
}

\date{\today}

\begin{abstract}

We use confocal microscopy to image colloidal gels formed from highly polydisperse particles. We suspend our polydisperse particles in a density matched solvent, and let the particles spontaneously aggregate through the van der Waals force. The particle size distribution $P(R)$ is roughly log-normal, with the largest particles more than 15 times the size of the smallest particles.  The pairing of nearest neighbor particles is consistent with a null hypothesis that pairings are made randomly, that is, any two particle sizes have a probability of being neighbors consistent with their proportionality in $P(R)$.  That being said, as expected, larger particles have more nearest neighbors than small ones.  This leads to an over-representation of large particles in tetrahedral structures where four particles are mutually nearest neighbors, showing that large particles help provide rigidity to the gel structure.  The tetrahedral structures also suggest that particles are able to rearrange during the gelation process, until their motion is stabilized by the multiple contacts with their neighbors.  We discuss the implications of how other size distributions $P(R)$ would affect the gel structure.

\end{abstract}

\maketitle

\section{\label{sec:intro}Introduction}

Colloids are suspensions of small solid particles suspended in a liquid.  ``Small'' means that the particle diameters range from $\sim$10~nm to $\sim$10~$\mu$m.  Thermal motion is relevant:  Brownian motion allows particles to diffuse.  Often precautions are taken to prevent the particles from sticking together \cite{jones02}.  If particles have attractive interactions, they can stick together in free-floating aggregates \cite{segre01prl,sciortino04,klix13}, or large tendrils that can span across the system \cite{dawson02,lu08,mohraz08,joshi14}.  The latter is a colloidal gel \cite{delgadobook16}.

In the 1980's and onward, several initial studies of colloidal gels focused on their fractal structure, analyzing two-dimensional images of flattened gels or using scattering techniques to measure the fractal dimension \cite{weitz_fractal_1984,weitz_limits_1985-2,tence_measurement_1986,dimon_structure_1986,liu_fractal_1990,piazza94,varadan01,bushell_techniques_2002}.  The fractal nature of these materials mean that the mechanical properties of a colloidal gel are significantly different from the pure liquid, even at low concentrations of the colloidal particles.  A fractal colloidal gel behaves like an elastic solid when the system-spanning tendrils can support stress over long ranges.  Later work used confocal microscopy to image the microstructure of colloidal gels in three dimensions \cite{dinsmore_direct_2002,varadan03,dinsmore06,dibble06,gao07,conrad08,lu08,ohtsuka08,pandey13,whitaker19,royall_direct_2008,mohraz08,dong22,tsurusawa_hierarchical_2023}.  This was complemented by simulations \cite{kolb_scaling_1983,dickinson13,zia_micro-mechanical_2014,puertas04,boromand17,padmanabhan18}.  Several groups observed local stable structures in the gels such as tetrahedra \cite{royall18,tsurusawa_hierarchical_2023,waheibi24}, triangular bipyramids \cite{royall_direct_2008,tsurusawa_hierarchical_2023}, and networks of these types of local clusters \cite{nabizadeh24}.  The stability of these structures is due both to their rigidity \cite{whitaker19} and the general fact that particles with more neighbors are more energetically stable \cite{royall_direct_2008,zia_micro-mechanical_2014}, that is, it is less likely for thermal fluctuations to detach highly connected particles from their position in the gel.  These various stable structures contribute to the mechanical properties of colloidal gels \cite{royall21}, tuning their mechanical properties such as rheological moduli, flowability, or resistance to gravitational effects \cite{buscall87,poon95,verhaegh97,starrs99,verhaegh99,dehoog01,starrs02,manley05collapse,huh07,lietorsantos10,bartlett12,teece14,razali17,padmanabhan18,Filiberti19}.

Much of this prior work studied gels composed of nominally monodisperse particles. While experimental colloidal particles are physical objects with some size variability, they are typically treated as identical \cite{sollich02, sollich_crystalline_2010}.  Some recent work has studied nominally bidisperse samples composed of small and large particles with a size ratio ranging from $1 : 8$ up to $1 : 24$ \cite{jiang22prl,li23}.  Given that the diffusive time scale for particles to diffuse their own size scales with radius as $R^3$, the larger species is effective non-Brownian in these samples \cite{li23,hunter12rpp}.  The inclusion of these large particles can distort the local structure of the gel composed of the smaller particles \cite{li23}, and can even result in rheological bistability \cite{jiang22prl}.

In our work, we study colloidal gels formed from a sample with a wide and continuous particle size distribution.  The largest particles are rare, but can be up to 6.5 times larger than the mean size and more than 15 times the size of the smallest particles. There is a continuous range of sizes and thus continuous variation of diffusive properties.  Larger particles have more surface area and can have more neighboring particles attached to their surface, but smaller particles diffuse more rapidly and thus could potentially find each other more easily in solution and/or find small corners between particles to fit into.  We wish to understand the rich structure formed in such colloidal gels, and therefore we do confocal microscopy of samples over a range of volume fractions from 0.01 to 0.45.  Our results show that such highly polydisperse gels are assembled fairly randomly -- all particles are essentially equally likely to be connected, in proportion to their prevalence in the particle size distribution.  Nonetheless, the larger particles have more nearest neighbors than smaller particles, and in fact act as sites of local rigidity in the colloidal gel structure.  Many particles contact three or more neighbors, suggesting that particles can rearrange to increase their number of contacts before the gel becomes fully immobile.

\section{\label{sec:methods}Experimental Methods}

\begin{figure}
    \centering
    \subfigure{\Large(a)}{\includegraphics[width=.92\columnwidth]{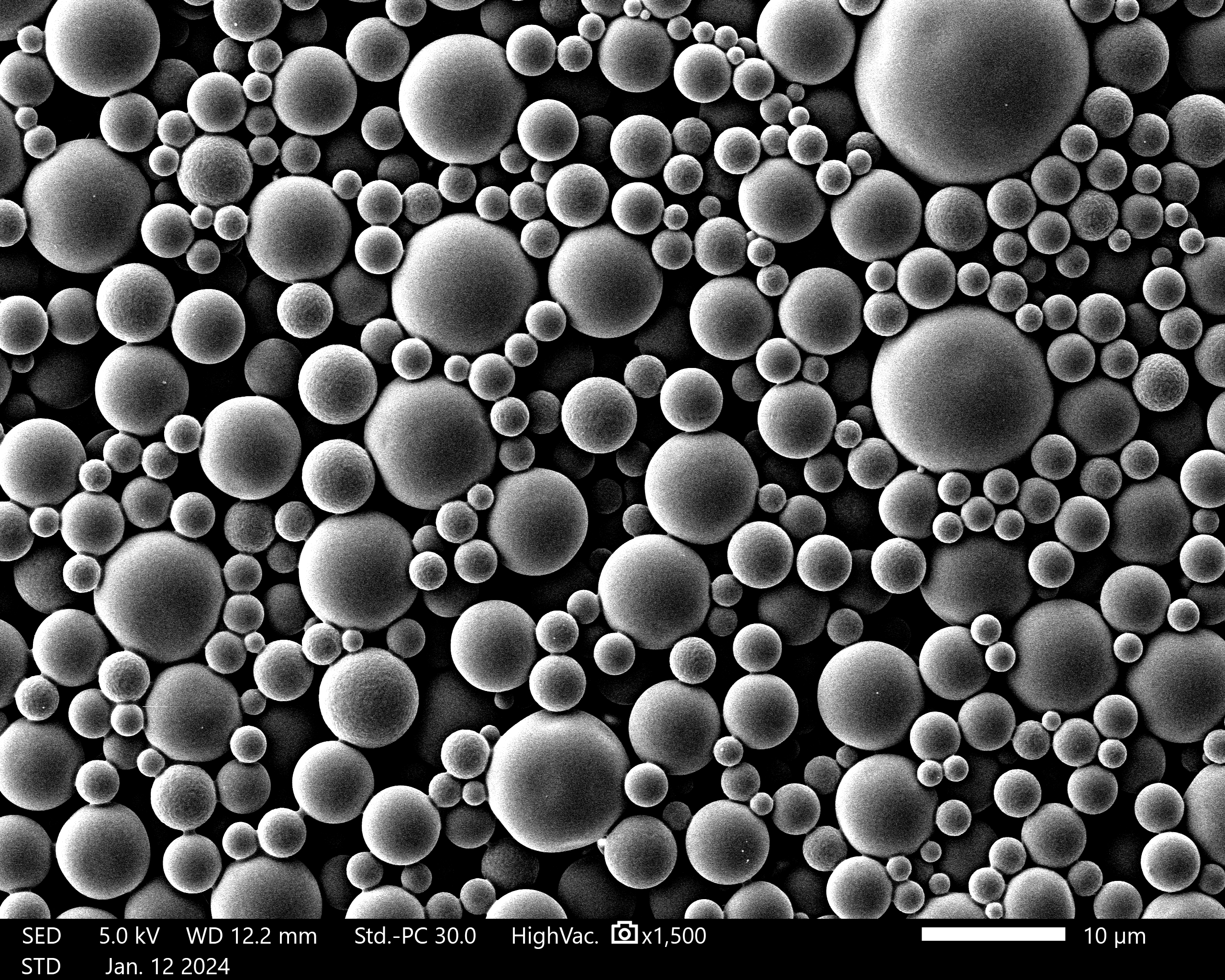}}
    \subfigure{\Large(b)}{\includegraphics[width=.92\columnwidth]{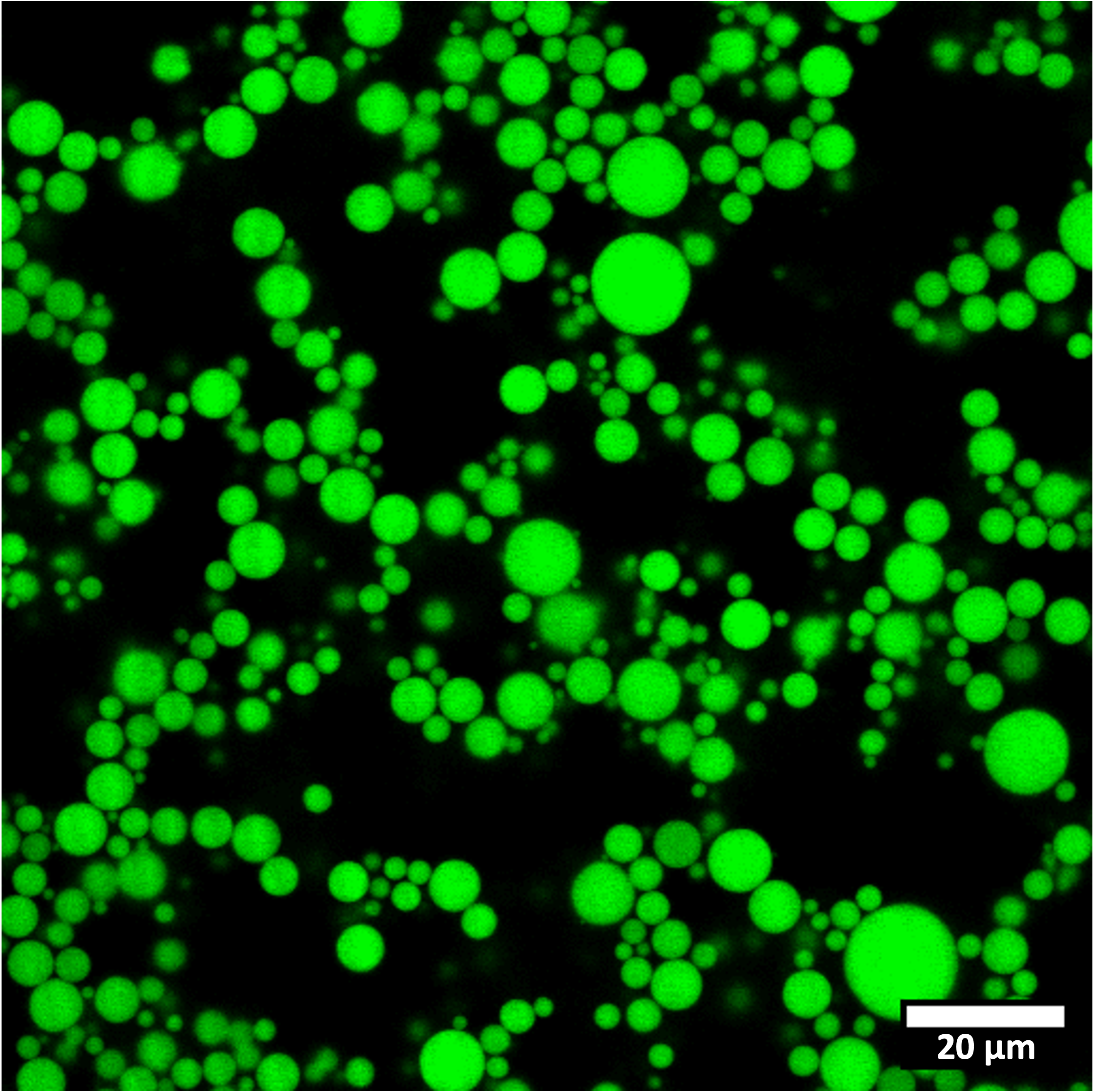}}
    \caption{(a) Scanning Electron Microscope image of colloidal PMMA particles (image courtesy of the Robert P. Apkarian Integrated Electron Microscopy Core at Emory University).  The scale bar is 10~$\mu${}m. (b) 2-dimensional confocal slice of fluorescently labelled colloidal particles.  The scale bar is 20~$\mu${}m and the volume fraction is $\phi=0.26$.}
    \label{fig:confocal}
\end{figure}

Our experiments use polydisperse spherical PMMA particles synthesized by A.B. Schofield (University of Edinburgh), via established methods \cite{antl86,campbell_fluorescent_2002}; images of these particles are shown in Fig.~\ref{fig:confocal}. As is clear from Fig.~\ref{fig:confocal}, the particles are highly polydisperse.  The particle sizes range from  1.0~$\mu${}m to 17.6~$\mu${}m.

Typically this synthesis method results in particles sterically stabilized by a thin layer of poly-12-hydroxystearic acid (PHSA), which prevents the particles from aggregating.  However, unlike much prior work with PMMA colloidal particles \cite{pusey86}, our colloidal particles are only imperfectly coated with PHSA.  This results in our particles aggregating automatically due to London-van der Waals forces.  For two spherical particles of radii $R_1$ and $R_2$, separated by a distance $r$, the interaction potential is given by
\begin{eqnarray}
\label{vdw}
    U(r) &=& - \frac{A}{6} \left[ \frac{2 R_1 R_2}{r^2 - (R_1 + R_2)^2} + \right.\\
    && \frac{2 R_1 R_2}{r^2 - (R_1 - R_2)^2} +
     \left. \ln \frac{r^2 - (R_1 + R_2)^2}{r^2 - (R_1 - R_2)^2} \right]\nonumber
\end{eqnarray}
where $A$ is the Hamaker constant \cite{mewisbook13,tadrosbook17}.  This formula was originally derived in 1937 by Hamaker \cite{hamaker1937,verweybook1948}.  Although we do not know the exact value for $A$ for our sample, it is typically $O(10^{-20}$~J) \cite{mewisbook13,larsonbook98,tadrosbook17,israelachvilibook}.  As a result of this large magnitude of $A$, the van der Waals force is quite strong, with the attractive energy at least ten times higher than $k_BT$ \cite{goodwin86,larsonbook98}, where $k_B$ is the Boltzmann constant and $T$ is the absolute temperature.  Thus, particles that touch are irreversibly bonded \cite{delgadobook16}.  However, given the $1/[r^2-(R_1+R_2)^2]$ term in Eqn.~\ref{vdw} which decays rapidly away from contact ($r=R_1+R_2$), these forces are quite short-ranged \cite{goodwin86,tadrosbook17}, and are smaller than $k_BT$ at surface-to-surface separations comparable to the particle sizes \cite{mewisbook13}.  The range of this force depends on the relative particle sizes, but not to a large extent.  As a concrete example, with $A = 2.5k_B T$ and particle sizes $R_1=R_2=1$~$\mu$m, $U(r) = k_B T$ at $r - R_1 - R_2=0.11$~$\mu$m.  If we increase $R_2$ to 100~$\mu$m, we have $U(r) = k_B T$ at $r - R_1 - R_2=0.22$~$\mu$m.  Note that Eqn.~\ref{vdw} is most accurate for small separations; at larger separations (more than 0.1~$\mu$m), the interaction energy decreases more strongly than the equation would suggest due to retardation effects, but there is no analytic formula describing this \cite{israelachvilibook,israelachvili1972}.  Nonetheless, qualitatively the behavior of the London-van der Waals force is as we have described:  much stronger than $k_BT$ for particles in contact, but weak for particles not in contact.

The confirmation that the van der Waals force is the relevant one in our experiments is that particle detachments from the gel structure are extremely rare in any of our observations, matching the known strength of this force, in contrast to other attractive forces such as depletion \cite{jones02} which in any case is not present in our sample as we have no depletant.  A key point is that while Eqn.~\ref{vdw} depends on both particle radii, qualitatively the van der Waals force is strong enough that the size-dependent properties do not matter for our experiments: all touching particle pairs will stick irreversibly, and the range of the attraction is only weakly dependent on particle sizes.

Our particles are in a solution of 51.5 vol\% decalin (Sigma Aldrich) and 48.5 vol\% tetrachloroethylene (Sigma Aldrich) \cite{kale_approaching_2023}.  This solvent mixture has several benefits.  First, the index of refraction of the colloids closely matches that of the background solvent, enabling optical microscopy so that the interior of the sample can be observed.  Second, this solvent mixture prevents ionic dissociation from the colloids into the background solvent, thus minimizing repulsive electrostatic interactions \cite{kale_approaching_2023}.  The only interparticle forces thus are the short-ranged van der Waals attraction and the solid core repulsion when particles are in contact.  Finally, the solvent mixture closely matches the density of the colloidal particles.  While the density of the particles has some variability, the particles require up to an hour at $400g$ in order to cause sedimentation. This close density-matching allows the colloidal gel structure to be stable during our observation time scale ($\sim$ minutes).

The samples are in chambers that have an interior volume of approximately $25 \times 10 \times 0.35$~mm$^3$.  Prior to adding the gel samples to the chambers, we vortex the stock jar of colloids for 3 minutes.  We next sonicate the jar for 30 minutes.  The samples are added to the microscope slides and then immediately put on a rotator which keeps the slides slowly rotating overnight, so as to further delay sedimentation.  The next day the samples are removed from the rotator and then imaged within an hour of removal.  We have separately confirmed with optical microscopy that immediately upon injection of the sample into the slides, the particles are mostly free monomers or dimers; the full gel structure thus forms overnight. 

We use confocal microscopy (Leica TCS SP8 MP) to image our colloids with a 63$\times$ 1.4 Numerical Aperture lens.  The images are typically $512 \times 512 \times 100$ voxels in size, with a voxel size of $0.27 \times 0.27 \times 0.30$~$\mu$m$^3$ or $0.34 \times 0.34 \times 0.30$~$\mu$m$^3$; both of these choices keep the voxel aspect ratio close to 1.  To avoid potential near-wall effects, our imaging volume is at least 35~$\mu${}m from the nearest wall. As noted above, the gels are imaged 12-20 hours after they are initially put into the sample chamber; by the time we image, particle motion is extremely arrested with most particles forming part of a gel. A typical two-dimensional confocal microscope image is shown in Fig.~\ref{fig:confocal}(b).

To determine the position and radii of our particles, we use the method of Penfold \textit{et al.} \cite{penfold_quantitative_2006}, along with additional techniques developed by Crocker \& Grier \cite{crocker_methods_1996}.  To look for three-dimensional (3D) spheres, we start by thresholding the image so that only voxels above a threshold are white, and all other voxels are black.  We then perform a distance transform on the binary image, producing a 3D Euclidean Distance Map (EDM).  In this map, the value of each voxel is the distance to the nearest black voxel of the binary image.  Potential spheres are identified by looking for local maxima in the EDM.  The precise center and radius are confirmed through a local spherical convolution of the raw image with the radius determined by the EDM. In dense gels especially, the EDM can have maxima in the bridge between contacting particles. We remedy this issue by looping through maxima from large to small radius and removing maxima that are found within the spherical shell determined by the local convolution. That is, if a small radius maxima is found within the shell of a larger radius maxima, then it is likely the larger radius that is the real particle. This method allows us to measure particle positions and radii with sub-voxel accuracy, although the exact radius measured depends on the choice of threshold used to form the binary image, leading to a systematic uncertainty for the radius of $\pm 0.2$~$\mu${}m based on the voxel size.  The uncertainty of our particle positions is half the voxel size, so $\pm 0.14$~$\mu$m.  The EDM method fails for particles about 4 pixels or less in radius, leading to a lower limit of identifiable particles of about 1.15~$\mu$m.

\begin{figure}
    \centering
    \includegraphics[width=\columnwidth,viewport=110 390 534 686]{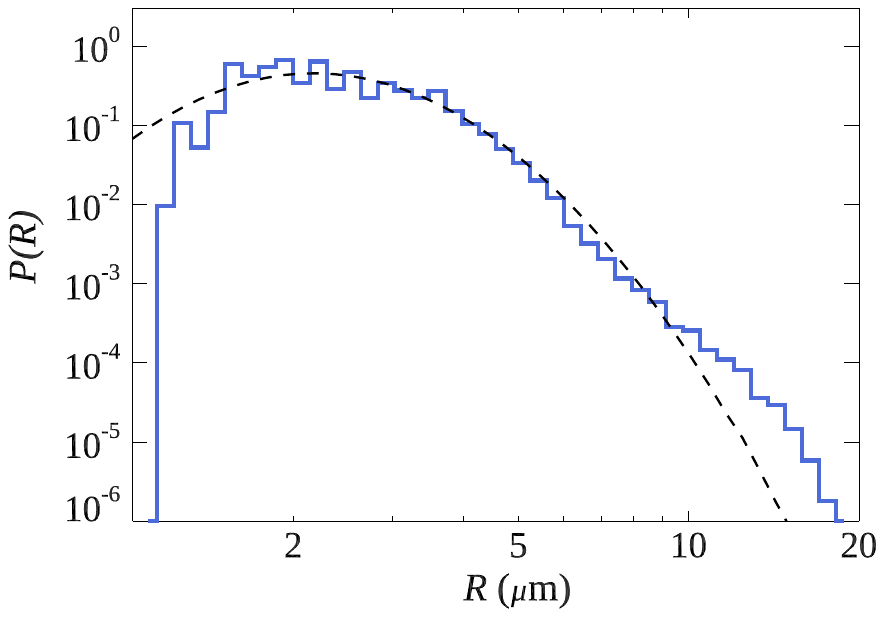}
    \caption{Measured probability distribution of the colloidal particle radii. The dashed black line corresponds to a log-normal fit (Eq.~\ref{lognorm} with $\sigma=0.378$). The mean size of the particles is $\mu=2.69$~$\mu${}m with a polydispersity of $\delta=37\%$. The total range of particle sizes goes from $1.15$~$\mu ${}m$ \leq r \leq 17.6$~$\mu ${}m.}
    \label{fig:rad_dist}
\end{figure}

Our particles are highly polydisperse and accordingly we show the particle size distribution in Fig.~\ref{fig:rad_dist}.  The mean size is $2.69$~$\mu${}m, the median size is $2.48$~$\mu${}m, and the polydispersity (standard deviation divided by the mean) is 0.37.  While 95\% of the particles have a radius $R \leq 4.5$~$\mu${}m, the largest (and rarest) particles have a radius of $R = 17.6$~$\mu${}m, 6.5 times larger than the mean size and more than 15 times larger than the smallest particles.  A common particle size distribution is log-normal \cite{kottler1950,middleton1970,sengupta1975,falco23}, given by:
\begin{equation}
    P(R) = \frac{1}{R \sigma \sqrt{2 \pi}} \exp \left( - \frac{(\ln R - \mu)^2}{2 \sigma^2}\right)
    \label{lognorm}
\end{equation}
with mean size $\mu$ and width $\sigma$.  A fit of our data to this $P(R)$ is shown in Fig.~\ref{fig:rad_dist} by the black dashed line, showing that our $P(R)$ is a bit like a truncated log-normal distribution.

We vary the volume fraction between $0.01 < \phi < 0.45$ by centrifuging the stock jar to a higher concentration before making the microscope slide, and then diluting the samples for making lower volume fraction slides. The volume fraction of our samples is measured locally by dividing the total volume of the spheres measured by the volume of the confocal stack \cite{poon12}; all volume fractions reported below are from local measurements.  Typically we take several images from different fields of view of each microscope slide.  In total, our data are 421 individual 3D images.  For all ranges of $\phi$ with width $\Delta \phi = 0.05$, we have at least 15 distinct images and more typically 25-40 distinct images.  The images for $\phi < 0.05$ are overrepresented in our data, with 106 images in this range.

\section{Results}
\label{sec:results}

\subsection{\label{sec:fractal}Fractal dimension}

The fractal dimension $D_f$ (or Hausforff dimension) is connected to the rugosity (or complexity) of a geometric object as different length scales of the object are probed \cite{mandelbrot_how_1967,mandelbrot_fractals_1989}.  Initially this idea was applied to self-similar objects, where viewing the object at different length scales results in similar images.  A different understanding of fractal dimension, however, has been successfully applied to characterize the microstructure of colloidal aggregates \cite{weitz_fractal_1984,tence_measurement_1986,bushell_techniques_2002,meakin_formation_1983,weitz_limits_1985-2,kolb_scaling_1983,meakin_effects_1988,farias_range_1996,piazza94,varadan01,hasmy_small-angle_1993,hasmy_small-angle_1994,liu_fractal_1990,bushell_fractal_1998,dimon_structure_1986,dinsmore_direct_2002,tolman_off-lattice_1989,krall_internal_1998}. The objective of this approach is less concerned with the complexity of the self-similar shape, but what might be termed the ``tenuousness" of the aggregate structure, with $D_f \rightarrow 1$ being a highly tenuous network and $D_f \rightarrow 3$ being a bulkier, denser network.  Hence, $D_f$ has been directly correlated with the elastic response of a gel:  the elastic modulus $G$ scales with volume fraction $\phi$ as $G \sim \phi^{1/(3-D_f)}$ \cite{krall_internal_1998,laurati_structure_2009}.  Beyond the mechanical properties such as elasticity, the fractal nature of a colloidal gel relates to the long-term stability of a gel \cite{manley05collapse,Filiberti19}, how solvent can flow through the network \cite{gelb19}, and provides insight into the gelation process \cite{lazzari_fractal-like_2016}.

To be precise, the fractal dimension of an aggregate relates the scaling of the aggregate mass, $M$, with its linear size, $s$, by a scaling relation $M \propto s^{D_f}$. Hence, a low fractal dimension aggregate spans a greater spatial scale than a high fractal dimension aggregate of the same mass. Prior work, both experimental and computational, has shown that the fractal dimension of aggregates of monodisperse particles typically hovers below $D_f\sim 2.0$ \cite{weitz_fractal_1984,tence_measurement_1986,bushell_fractal_1998}. 

\begin{figure}
    \centering
    \includegraphics[width=\columnwidth,viewport=114 376 526 687]{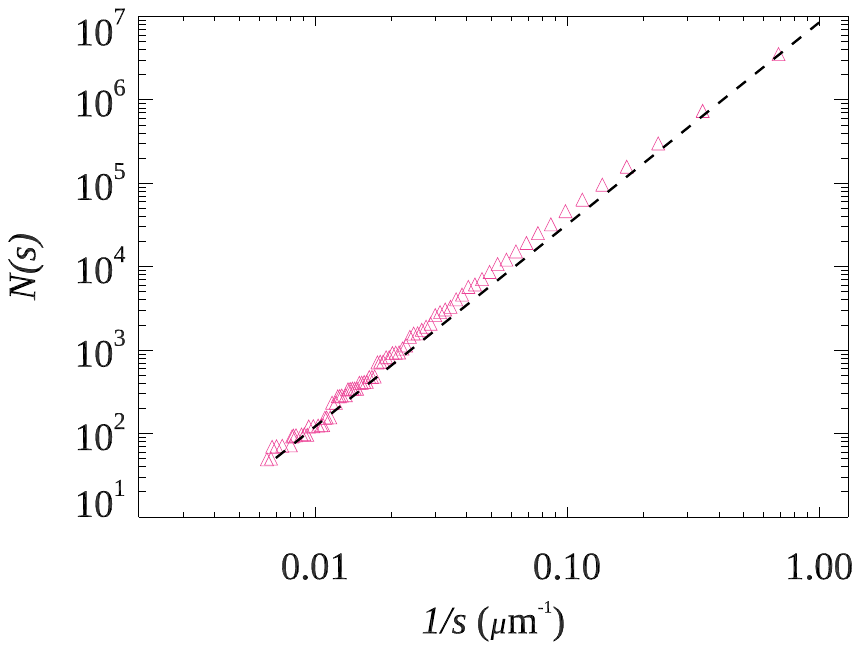}
    \caption{Graph of box counting method to estimate $D_f$ for one image stack of a gel at $\phi \sim 0.15$. To simplify the calculation of $D_f$, the aspect ratio of the voxel is made uniform. The slope of inverse box size ($s$) against the number of boxes with a bright voxel ($N(s)$) gives the fractal dimension of the aggregate. The dotted black line fits to the equation $N(s)=s^{-D_f}$ where $D_f=2.43$. We repeat this measurement for three other image stacks and report a mean value of $D_f=2.5 \pm 0.1$.}
    \label{fig:fracdim}
\end{figure}

In our experiment, we use an intermediate volume fraction gel ($\phi \sim 0.15$) and measure $D_f$ through a ``box-counting" method. This method discretizes a 3-dimensional binarized confocal image into cubes of edge lengths $s$ and counts how many cubes contain at least one bright voxel.  By starting at the smallest box size of one voxel and incrementing up to the size of the image stack, we plot the box count against the size of the box in Fig.~\ref{fig:fracdim}.  Fitting to $N \sim s^{-D_f}$ yields a fractal dimension of $D_f=2.5 \pm 0.1$ for our polydisperse colloidal gels, with the uncertainty due to the thresholding choice and also reflecting an average across several different gels; see details in the Appendix.

Much prior work has studied the fractal dimension of various colloidal gels, and it is known that the value of $D_f$ depends on several factors such as the attractive energy of sticking in experiments \cite{weitz_limits_1985-2,liu_fractal_1990,ohtsuka08}, the aggregation kinetics \cite{bushell_techniques_2002}, the attraction range \cite{mangal24}, and/or the simulation algorithm \cite{eggersdorfer_structure_2012,tomchuk_modeling_2020-1}.  Bushell {\it et al.} noted that $D_f$ can range from 1.5 to 2.5 depending on the details of gel formation \cite{bushell_techniques_2002}.  Given the variety of these factors that influence the fractal dimension, we do not expect our $D_f = 2.4$ to match any particular result. Nevertheless, a comparison with previous work is instructive as to some factors that may be relevant for our particular colloidal gel formation.  Some of this prior work specifically considered the role of particle polydispersity. Intuitively, if fractal dimension is a measure of the spatial scaling of an aggregate, then a polydisperse distribution of spherical particles might be a way to embed ``extra" scaling and thus modify the overall fractal dimension of an aggregate. An early experiment found that $D_f = 1.9 \pm 0.1$ was similar for colloidal aggregates formed from monodisperse and polydisperse particles \cite{tence_measurement_1986}.  A contemporary simulation found $D_f = 1.78 \pm 0.03$ with a similar lack of difference between monodisperse and polydisperse particles \cite{bushell_fractal_1998}.  Later computational work found that the fractal dimension monotonically decreases with increasing polydispersity for two common algorithms, Diffusion Limited Cluster Aggregation (DLCA) and Diffusion Limited Aggregation (DLA) \cite{eggersdorfer_structure_2012,chou_effects_1996}.  Eggersdorfer \& Pratsinis \cite{eggersdorfer_structure_2012} argue that by only testing Gaussian distributions with a maximum geometric standard deviation of $\sigma_{max}=1.5$, these earlier works (Refs.~\cite{tence_measurement_1986,bushell_fractal_1998}) failed to probe a large enough polydispersity to notice the effect.  Our geometric standard deviation is $\sigma = 1.41$, so not in the extreme polydispersity limit considered by Ref.~\cite{eggersdorfer_structure_2012}.  The fractal dimensions measured by Eggersdorfer \& Pratsinis range from 2.25 to 1.48 as $\sigma$ increases from 1.0 to 3.0, with the details also dependent on the algorithm (DLCA or DLA).  Using DLCA and $\sigma = 1.0$ recovered $D_f = 1.79 \pm 0.03$ in agreement with Ref.~\cite{bushell_fractal_1998}, which drops slightly to $D_f = 1.77 \pm 0.03$ for $\sigma = 1.45$.  Our $D_f = 2.5$ is significantly larger.

We believe that the likely reason for our higher fractal dimension is that our colloidal particles, while attracted to each other, still have the possibility of rearranging. Our reasoning stems from experimental work by Liu \textit{et al.} which showed the influence of the interparticle attraction strength on the fractal dimension of colloidal gold aggregates \cite{liu_fractal_1990}. In their study, they controlled the attraction strength by varying the concentration of added surfactant. At low to moderate surfactant concentration, surfactants reduce the electrostatic repulsion between the gold particles by coating their surface and adsorbing residual surface charges, thus increasing the interparticle attraction energy in proportion to the concentration of added surfactant. At a low surfactant concentration/low attraction energy, they were able to produce dense colloidal aggregates with $D_f=2.7$. With the addition of extra surfactant/raising the attraction energy, they showed that the aggregates open into sparser structures with $D_f=1.7$ \cite{liu_fractal_1990}.  This phenomenon of low energy/dense aggregates is likely due to the inability of single particle bonds to sequester particles into a gel, thus allowing colloids to perform a constrained diffusion along the surface of its neighbor until they settle into a more rigid conformation with more bonds \cite{ohtsuka08,zia_micro-mechanical_2014,puertas04,mangal24}.  This secondary, post-sticking diffusion has been studied in Meakin \& Julien where the authors show that restructuring can carry an aggregate from $D_f=1.8$ all the way to $D_f=2.2$ \cite{meakin_effects_1988}.  We believe this restructuring likely occurs in our colloidal gel structures before observation, allowing our fractal dimension to rise to the observed $D_f = 2.5 \pm 0.1$.

\subsection{\label{sec:level4}Contact network}

The contact number $Z$ is an important metric for characterizing a gel network, as it gives direct insight into the connectivity of the microstructure.  Whereas the fractal dimension characterizes the gel over a range of length scales larger than the particle size, the contact number is a local measure at the particle scale, thus providing complementary information.  For monodisperse particles the contact number will depend on the strength of the attractive forces and the overall volume fraction, but is geometrically restricted to 12 contacts.  Dilute samples with low attractive energy might just have occasional dimers or trimers and the mean contact number could be less than 1.  High volume fraction samples such as attractive colloidal glasses \cite{pham02,pham04,royall2012quench} could have contact numbers approaching 10-12.  One early study used confocal microscopy to measure contact numbers for a monodisperse colloidal gel, finding $Z$ ranged between $2.7 < \langle Z \rangle <4.0$ depending on the strength of attraction \cite{dinsmore_direct_2002}. Later work found wider ranges of contact numbers, depending on the details of the particle interaction \cite{dibble06,lee08,ohtsuka08,rice12,dong22,nabizadeh24}.  Computational and experimental work has revealed the role of structural evolution in the mechanical properties of colloidal gels \cite{zia_micro-mechanical_2014, tsurusawa_hierarchical_2023,datta2025,zhang2013, bonacci_contact_2020,dong22,ohtsuka2008local,royall18}. Many different features of the coarsening process have been highlighted (e.g. relaxation differences between colloids at the surface of a strand vs. in the bulk of a strand; hierarchical assembly of various structural motifs; contact stiffening of solid-solid bonds), but the end effect on the aggregate is similar: a moderate compaction of the contact network with an increase in the average contact number as the gel evolves.

For polydisperse gels, the theoretical maximum of contacts is highly dependent on the shape of the particle size distribution in addition to these other factors. In particular, the ``granocentric" model developed by Clusel \textit{et al.} establishes analytic relationships between particular size distributions and expectation values for the number of contacts \cite{clusel_granocentric_2009,corwin_model_2010}, but only for the high $\phi$ regime of random close packing.

\begin{figure}
    \centering
    \includegraphics[width=\columnwidth,viewport=73 373 288 623]{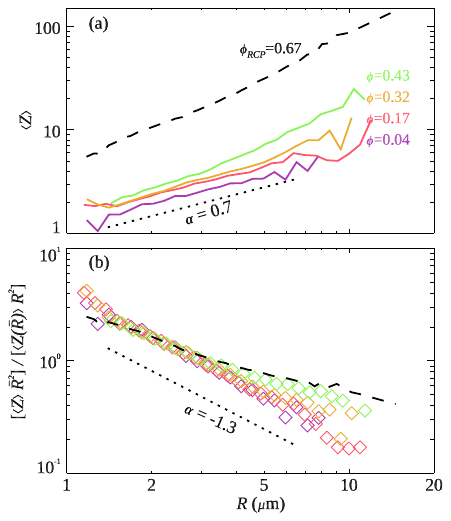}
    \caption{(a) The average contact number $\langle Z \rangle$ as a function of the radius of the particle $R$.  The volume fraction $\phi$ increases monotonically from bottom to top as labeled. The dashed black curve represents data gathered from RCP simulation which pack at $\phi_{RCP}=0.67$. (b) The data from (a) is replotted as $[  \langle Z \rangle \bar{R}^2] / [ \langle Z(\bar{R} \rangle R^2]$, using the mean particle size $\bar{R} = 2.69$~$\mu${}m.  This normalizes the data by the contact number corresponding to $\bar{R}$, as well as dividing by $R^2$ to remove the influence of the surface area of each particle.  For both (a) and (b), the dotted lines follow a power law with slopes $\alpha$ as indicated.
    }
    \label{fig:contact}
\end{figure}

In Fig.~\ref{fig:contact}, we plot the average number of contacts $\langle Z \rangle$ for a particle of radius $R$; the different curves indicate different volume fractions $\phi$. Contacts are defined as particles whose separation $d$ is less than or equal to $r_1 + r_2 + 2\delta r$, where $\delta r=0.2$~$\mu$m is the uncertainty in the radius as determined by particle tracking; this is also a distance where the van der Waals interaction decreases to less than $k_BT$, as described previously.   Several qualitative observations are straightforward.  First, larger particles have more contacts; a result in line with their greater surface area. Second, at low volume fractions, particles still form contacts; a feature indicative of a colloidal gel.  Even for the smallest particles and the smallest volume fraction, $\langle Z \rangle > 1$.  Third, at higher volume fractions, particles have more contacts, an unsurprising result \cite{waheibi24}.

As seen in Fig.~\ref{fig:contact}, the relation between $\langle{}Z\rangle$ and $R$ appears to grow slightly faster than a power law. The curvature also appears to increase as a function of $\phi$.  We note that there is no reason to expect a power-law relation or any other particular relation, given that our particle size distribution does not have a simple analytic form (see Fig.~\ref{fig:rad_dist}).  To better understand the relation between $\langle{}Z\rangle$ and $R$, we examine simulated random close packing (RCP) structures based on our measured particle size distribution given by Fig.~\ref{fig:rad_dist}.  We use the simulation methods described in Desmond \& Weeks \cite{desmond_influence_2014} as modified in Meer \& Weeks \cite{meer24} to generate close packed configurations with 800 to 3200 particles, finding $\phi_{\rm RCP} = 0.675 \pm 0.004$.  In Fig.~\ref{fig:contact}, we add the computational results as the top black dashed line.  Our experimental data for highest $\phi$ mimic the shape of this curve but with lower values.

To partially collapse the data, we divide the contact number by the number of contacts at the mean radius $\bar{R}=2.7$ and replot the data in Fig.~\ref{fig:contact}(b), confirming the agreement between the $R$ dependence of the high $\phi$ experimental data and the RCP data.  We additionally have normalized the data by multiplying by $\bar{R}^2/R^2$ to remove the surface area dependence.  The corresponding decrease shows that while larger particles have more neighbors, the number of neighbors is fewer than would be expected if the surface area were the sole determining factor.  In Fig.~\ref{fig:contact}(b),
the lower $\phi$ experimental curves deviate for large $R$ from the RCP data, showing that the largest particles have fewer contacts than might be expected in these low volume fraction colloidal gels.  Nonetheless, the qualitative agreement between the experimental and computational data is intriguing given that the computational results treat the particles as purely repulsive hard spheres, and do not have any physical dynamics or attractive interactions considered.  This strongly suggests that the experimental particles pair up randomly.

\begin{figure}
    \centering
    \includegraphics[width=0.9\columnwidth,viewport=117 387 534 686]{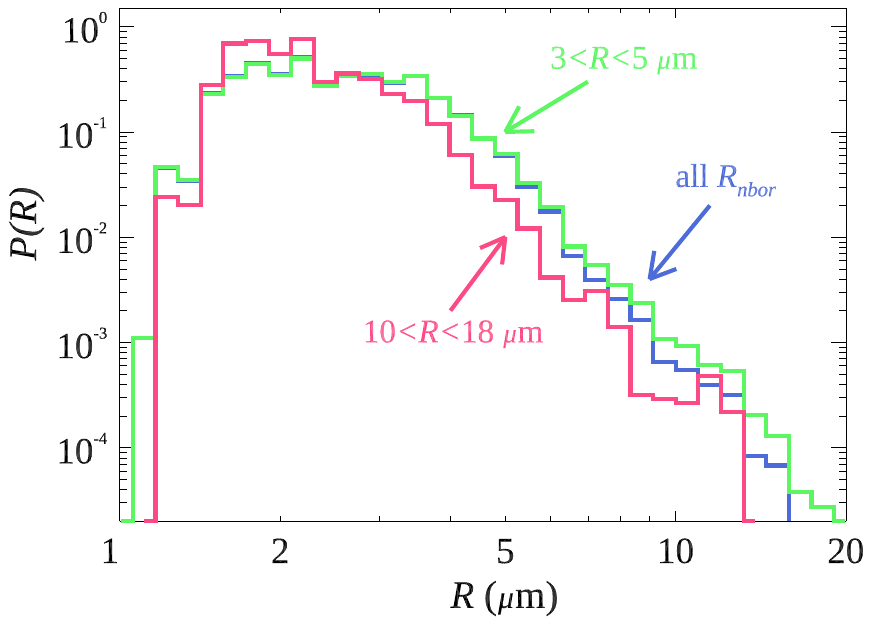}
    \caption{Histograms of the particle size distribution given that it is neighbors with a particle of size $3.0<R<5.0$~$\mu$m (green) and $10.0<R<18.0$~$\mu$m (pink) for the experimental data. The blue distribution is the ``contact former" distribution, {\it i.e.}, the list of particle sizes that form contacts.}
    \label{fig:nbor_comp}
\end{figure}

We do one additional test to verify that particles pair up randomly. Figure~\ref{fig:nbor_comp} shows the probability distribution of contact particle sizes, conditional on the size of the particle.  The curves are for contacts of small particles ($3<R<5 \mu$m), contacts of large particles ($10<R<18 \mu$m), and all contacts.  That is, for ``all contacts,'' we list all particle pairs, extract their radii, and take a histogram of the data.  A particle with $Z$ contacts will have its radius listed $Z$ times.  The data of Fig.~\ref{fig:nbor_comp} show that small and large particles have nearly equivalent probability distributions for the size of their contacts.  The distribution for the largest particles (pink curve) is shifted slightly to the left:  the mean size of neighbors for small particles is $3.0 \pm 1.2$~$\mu$m, and for large particles it is $2.5 \pm 0.9$~$\mu$m (mean $\pm$ standard deviation.  Large particles are slightly more likely to have smaller neighbors, and vice versa for small particles, but the difference is not dramatic.

\subsection{\label{sec:tets}Tetrahedral structure}

\begin{figure}[bthp]
    \centering
    \subfigure(a){\includegraphics[width=0.9\columnwidth]{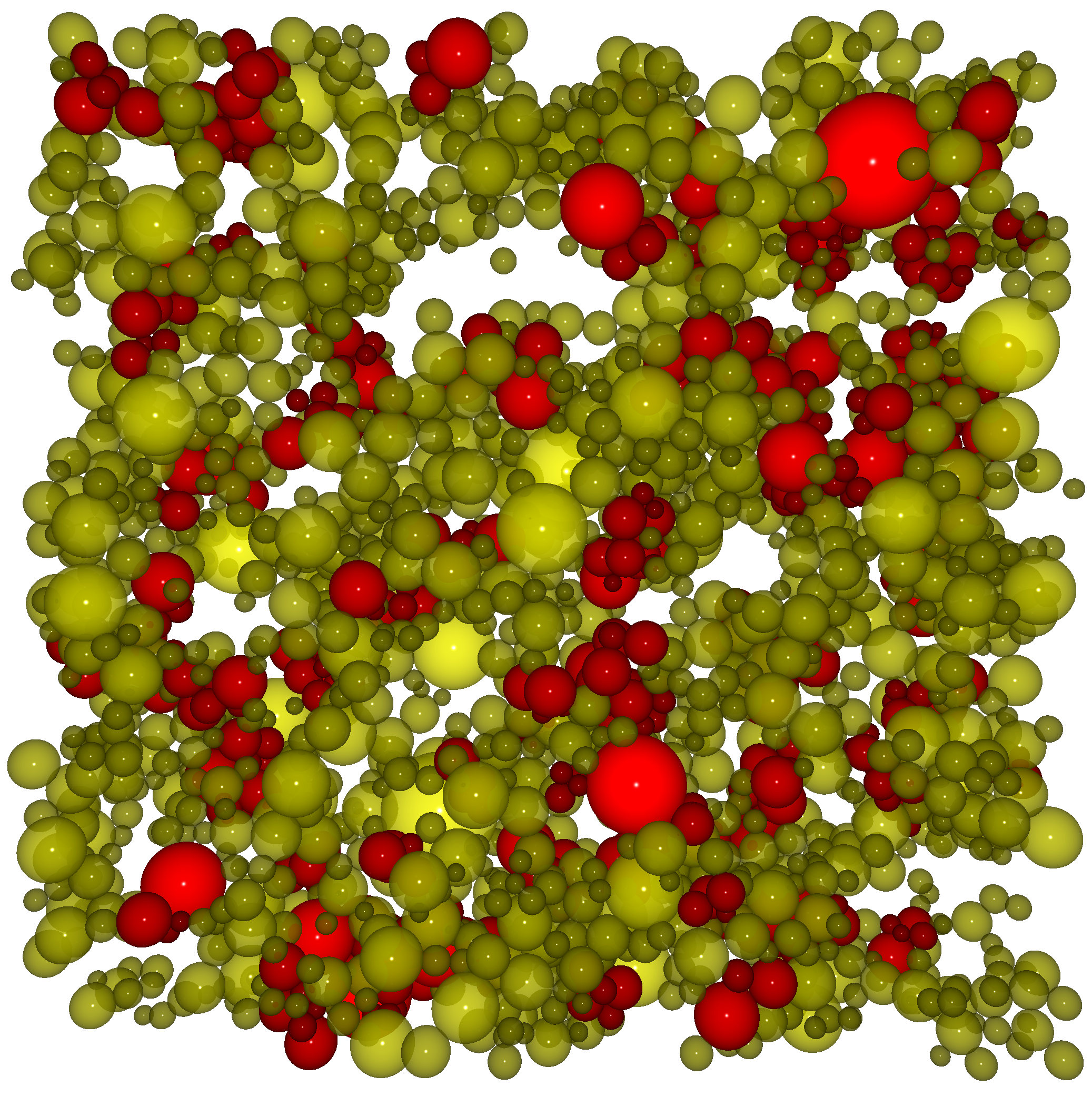}}\\
    \subfigure(b){\includegraphics[width=0.42\columnwidth]{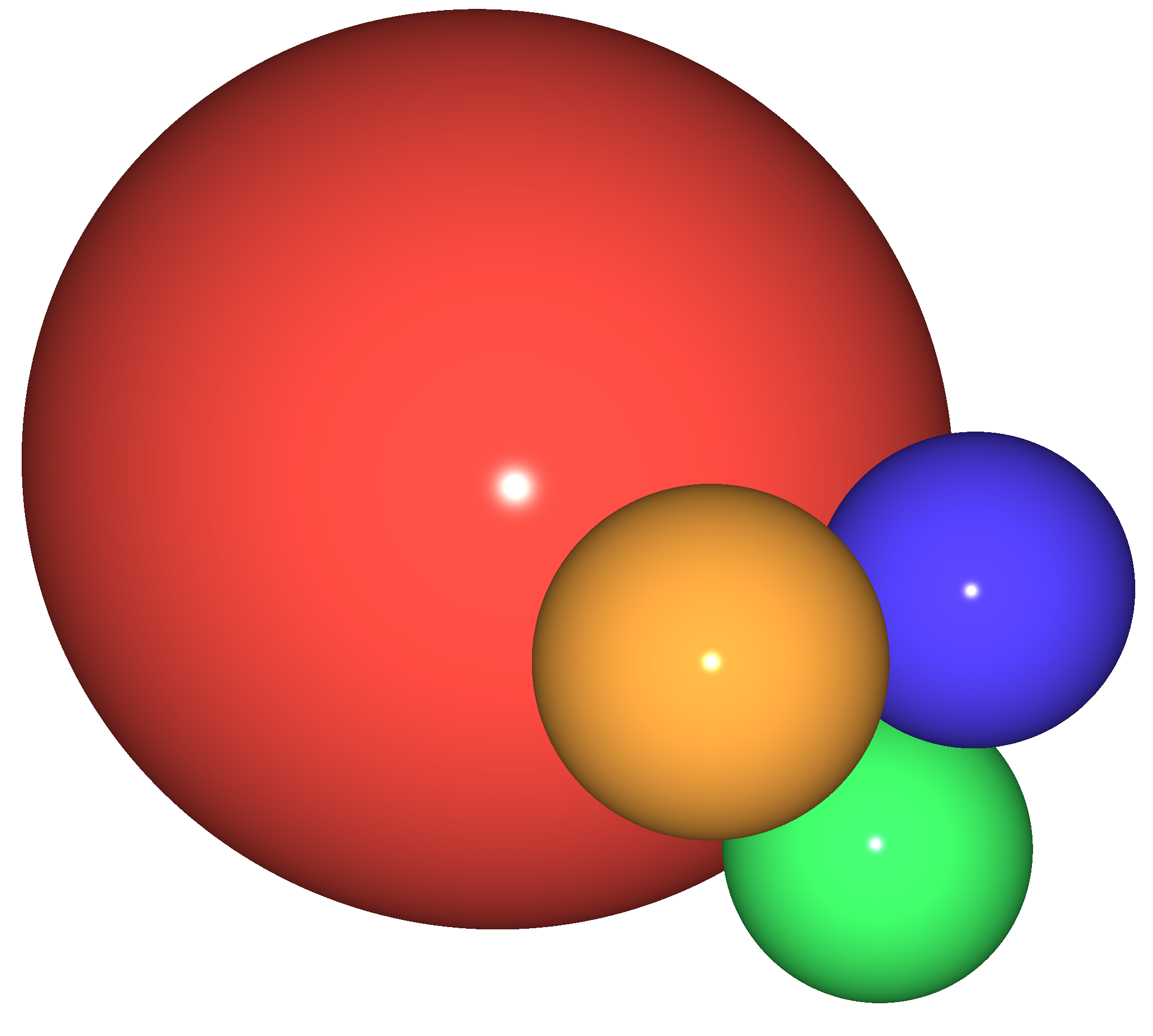}}
    \subfigure(c){\includegraphics[width=0.42\columnwidth]{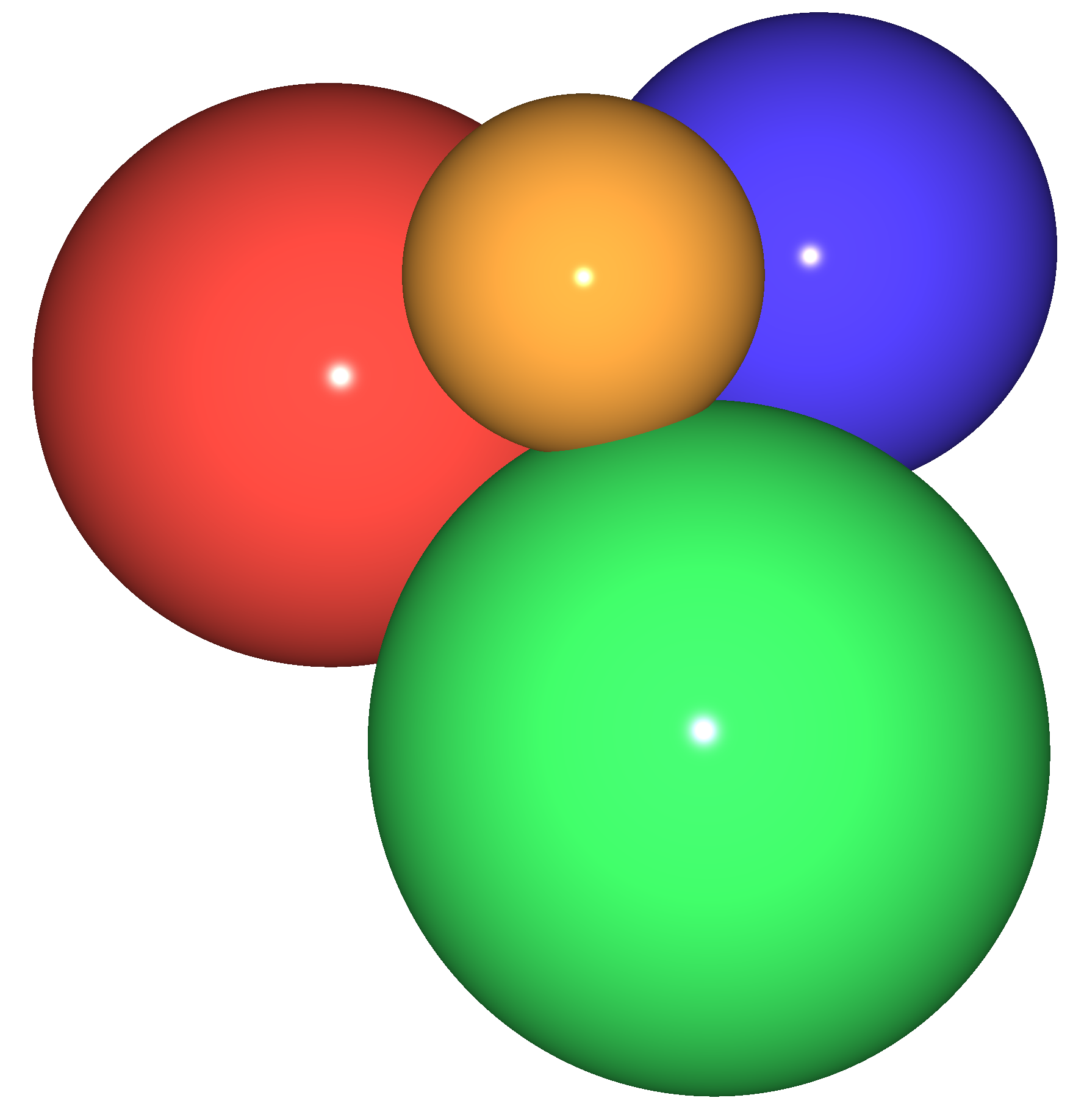}}
    \caption{(a) A 3D rendered image of a colloidal gel. Particles in dark red are engaged in a tetrahedron with the shading determined by particle size, and the lighter yellow particles are other particles in the gel. The width of the image is 138~$\mu${}m and $\phi=0.27$.  (b-c) 3D rendered tetrahedra from our sample. In (b), the largest particle has a radius $R=9.1$~$\mu${}m and the asymmetry parameter for this tetrahedron is $\tau=1.42$. In (c), the largest particle has a radius $R=3.6$~$\mu${}m, and $\tau = 1.51$, indicating a slightly more symmetric tetrahedron than the one shown in (b). }
    \label{fig:tet3d}
\end{figure}

Our final structural characterization is an examination of tetrahedral particle configurations, to complement the fractal dimension (large scale) and coordination number (single particle).  Tetrahedral structure examines how four particles are mutually neighbors, and relates to local rigidity.
The onset of an elastic response well below the isostatic percolation determined by Maxwell counting is a peculiarity of colloidal gels \cite{tsurusawa_direct_2019,maxwell_l_1864,cui19,zhang19prl}.  Microstructurally, it has been shown that for some gels, the dynamical arrest brought on by gelation and the subsequent elasticity of the gel correspond to the formation of rigid tetrahedral structures that percolate as the gel ages \cite{tsurusawa_hierarchical_2023,richard_coupling_2018,royall_direct_2008,haxton_crystallization_2015,nabizadeh24}.  The concept is that four particles that are mutually nearest neighbors form a tetrahedron, and this structure cannot be flexed without breaking a contact \cite{royall18,waheibi24}.  If the contacts are fairly strong (compared to $k_B T$), this imparts rigidity to the local structure \cite{maxwell_l_1864}.  If there is a percolation of such tetrahedral structures, the gel will be macroscopically rigid.  To be precise, in Ref.~\cite{tsurusawa_hierarchical_2023} Tsurusawa \& Tanaka cite the pentagonal bipyramid, not strictly the tetrahedron, as the percolating, rigidity enhancing structure. They do so because the pentagonal bipyramid has a five-fold symmetry that prevents crystallization, thus offering a route to understanding rigidity outside the traditional notions of local glassy patches or crystallization (the icosahedron has also been suggested for its similar five-fold symmetry). For a detailed explanation of crystalline frustration in colloidal gels, see Royall \textit{et al.} \cite{royall_direct_2008}.  

For our polydisperse gels, where crystallization is impossible, the tetrahedron is a simple and useful structure that can lead to rigidity.  Indeed, our colloidal gels have tetrahedral clusters of particles.  Figure \ref{fig:tet3d}(a) shows a 3-dimensional rendering of one sample, with identified tetrahedra colored in red. As shown in Fig.~\ref{fig:tet3d}(b,c), a tetrahedron occurs when four particles are simultaneously in contact with each other, meaning that tetrahedra can be identified by looking for sets of common entries in the contact array. 

\begin{figure}
    \centering \includegraphics[width=\columnwidth,viewport=69 369 239 659]{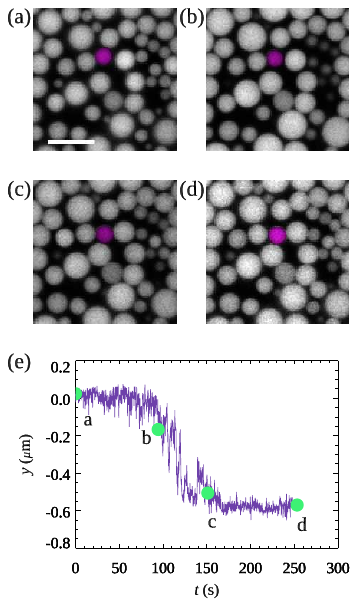}
    \caption{(a-d):  Images showing a highlighted particle which, over 250~s, moves from a position where it touches two other particles, into a more stable position where it touches three other particles.  The scale bar indicates 5~$\mu$m, and the highlighted particle is $R=1.1$~$\mu$m in radius.  The sample was photobleaching during the observation, so the contrast was adjusted by hand during acquisition of this movie.  Thus (d) is slightly noisier despite also being slightly brighter.  (e) The $y$ coordinate of the highlighted particle as a function of time, with the points corresponding to panels (a-d) as indicated. }
    \label{fig:traj}
\end{figure}

\begin{figure}
    \centering    \includegraphics[width=\columnwidth,viewport=65 323 366 724]{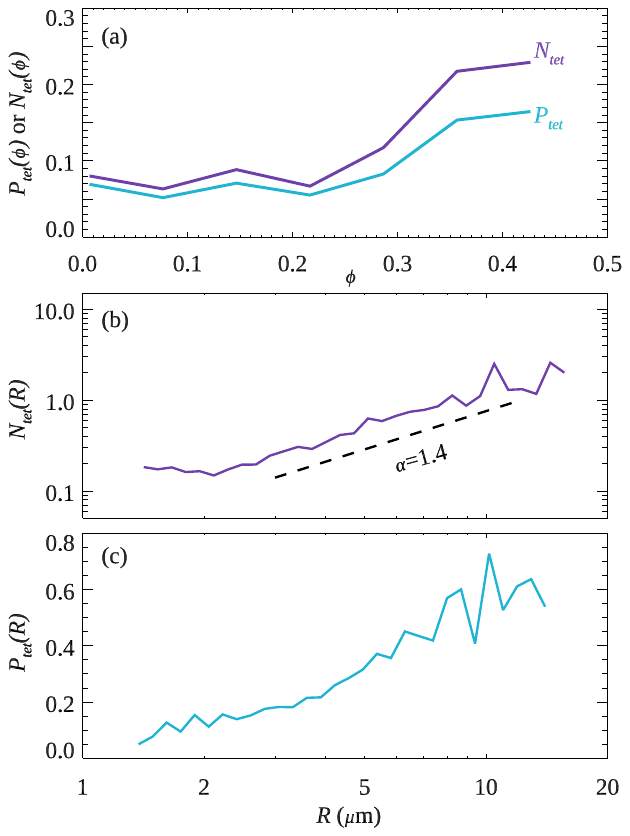}
    \caption{(a) The probability $P_{\rm tet}(\phi)$ that a particle is in a tetrahedron as a function of volume fraction $\phi$, and also the number $N_{\rm tet}$ of tetrahedra on average that a particle participates in.  At low $\phi$, $P_{\rm tet} \sim N_{\rm tet}$, suggesting that if a particle is in a tetrahedron, it is unlikely to also be in a second tetrahedron.  The separation of the two curves at higher $\phi$ shows that more particles participate in multiple tetrahedra.  (b) $N_{\rm tet}(R)$ as a function of particle radius $R$ averaged over samples with $\phi > 0.35$.  The plot has logarithmic axes and the dashed line indicates power-law growth with an exponent $\alpha=1.4$.  (c) $P_{\rm tet}(R)$ as a function of $R$ for the same data as (b).}
    \label{fig:pntet}
\end{figure}

The probability of a particle being found in at least one tetrahedron is shown by the lower curve in Fig.~\ref{fig:pntet}(a), and the number of tetrahedra the average particle participates in is shown by the upper curve.  Unsurprisingly, both rise with larger volume fraction, although intriguingly even at small volume fraction tetrahedra are present.  Note that the data shown in Fig.~\ref{fig:pntet} are based on particles within the imaging volume; it is likely that the results are a slight underestimate, as particles may be part of tetrahedra that extend outside the imaging volume. 

The existence of tetrahedral structures implies that particles have some ability to rearrange even after initially sticking together.  If particles that initially touch form an irreversible bond preventing rearrangements, then many particles would only have one or two nearest neighbors.  On the other hand, if the bond energy is not too large (a few $k_B T$) then particles with one nearest neighbor would have some ability to detach and reattach.  The implication is that the more neighbors a particle is bonded to, the more stable its position would be \cite{royall_direct_2008,zia_micro-mechanical_2014,puertas04,mangal24}.  This would be equally true if a particle can ``roll'' or ``slide'' while remaining attached to its neighbor.  Indeed, the London-van der Waals attraction is a central force, so there is no bond rigidity.  A particle touching one other particle can roll and/or slide freely on its surface; a particle touching two other particles can roll and/or slide in the groove formed by its two neighbors.  Only a particle touching three other particles is fully constrained from moving -- and if those three other particles are also mutually nearest neighbors, they have formed a tetrahedron.  Thus, the existence of tetrahedral structures in our gels, indeed a fair number of those structures, suggests that our particles are able to rearrange somewhat as the gel forms (as we had previously suggested in Sec.~\ref{sec:fractal}).  A two-dimensional movie taken at an early stage of gel formation (within 30 minutes after stirring the sample) has these sorts of rearrangement events, one of which is pictured in Fig.~\ref{fig:traj}.  In this example, the highlighted particle moves from a position where it touches two neighbors into a more stable location where it touches three neighbors, upon which it ceases motion.  Note that during the period from (b) to (c), the particle is much more mobile, as indicated by the fluctuations in the $y$ position shown in Fig.~\ref{fig:traj}(e).  We have observed other particles moving on the surface of a particle to which they are bonded.

There is additionally a particle size dependence for the tetrahedra.  To illustrate this, in Fig.~\ref{fig:pntet}(b,c) we plot the number of tetrahedra $N_{\rm tet}(R)$ and probability of being in a tetrahedron $P_{\rm tet}(R)$ as a function of particle radius (considering only the data for which $\phi > 0.35$).  Both rise dramatically with particle radius.  As Fig.~\ref{fig:contact}(a) shows, large particles have more contacting neighbors.  More neighbors gives more chances that those neighbors are themselves mutually in contact, forming a tetrahedron. Although Fig.~\ref{fig:pntet}(b) shows a power law relation, as noted before for the contact number relation shown in Fig.~\ref{fig:contact}, there is no particular reason to expect a power law relation for these data.

\begin{figure}[bthp]
    \centering
    \includegraphics[width=0.9\columnwidth,viewport=118 393 535 688]{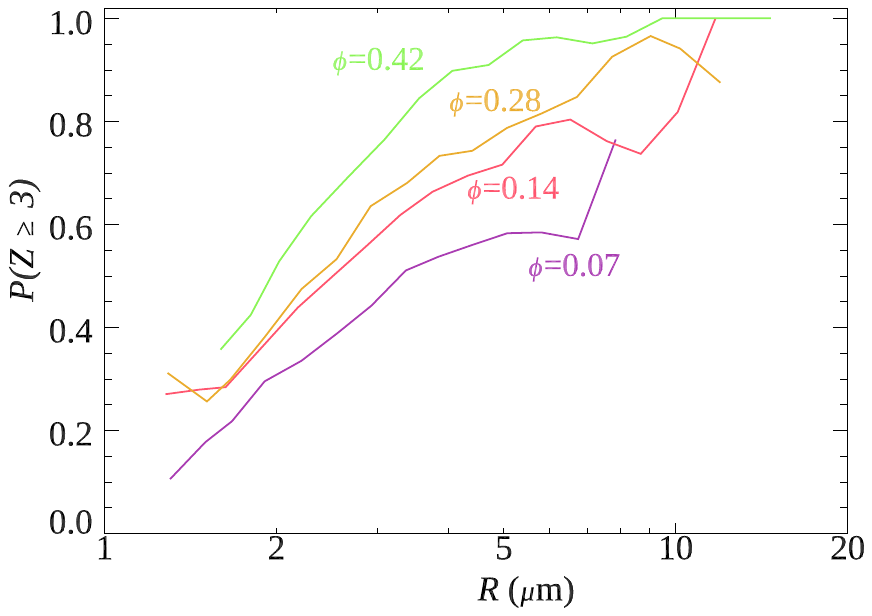}
    \caption{The probability of a particle of radius $R$ to have a contact number $Z \ge 3$, that is, to have at least three neighbors.  Different curves are from different volume fractions as indicated. }
    \label{fig:pz3}
\end{figure}

It is perhaps surprising that small particles ($R < 2$~$\mu$m) have any chance of being in a tetrahedron, given that Fig.~\ref{fig:contact}(a) shows they have $Z = 1 - 2$ neighbors on average.  However, the mean values shown in Fig.~\ref{fig:contact}(a) do not reflect the outliers.  Figure~\ref{fig:pz3} shows the probability of particles with radius $R$ to have at least three neighbors, which is about 10-30\% depending on the volume fraction.  Thus, these outlier particles have some chance to be in tetrahedra.  As expected, Fig.~\ref{fig:pz3} shows that the largest particles are much more likely to have at least three neighbors.

\begin{figure}
    \centering
    \subfigure{\Large(a)}{\includegraphics[width=0.4\columnwidth]{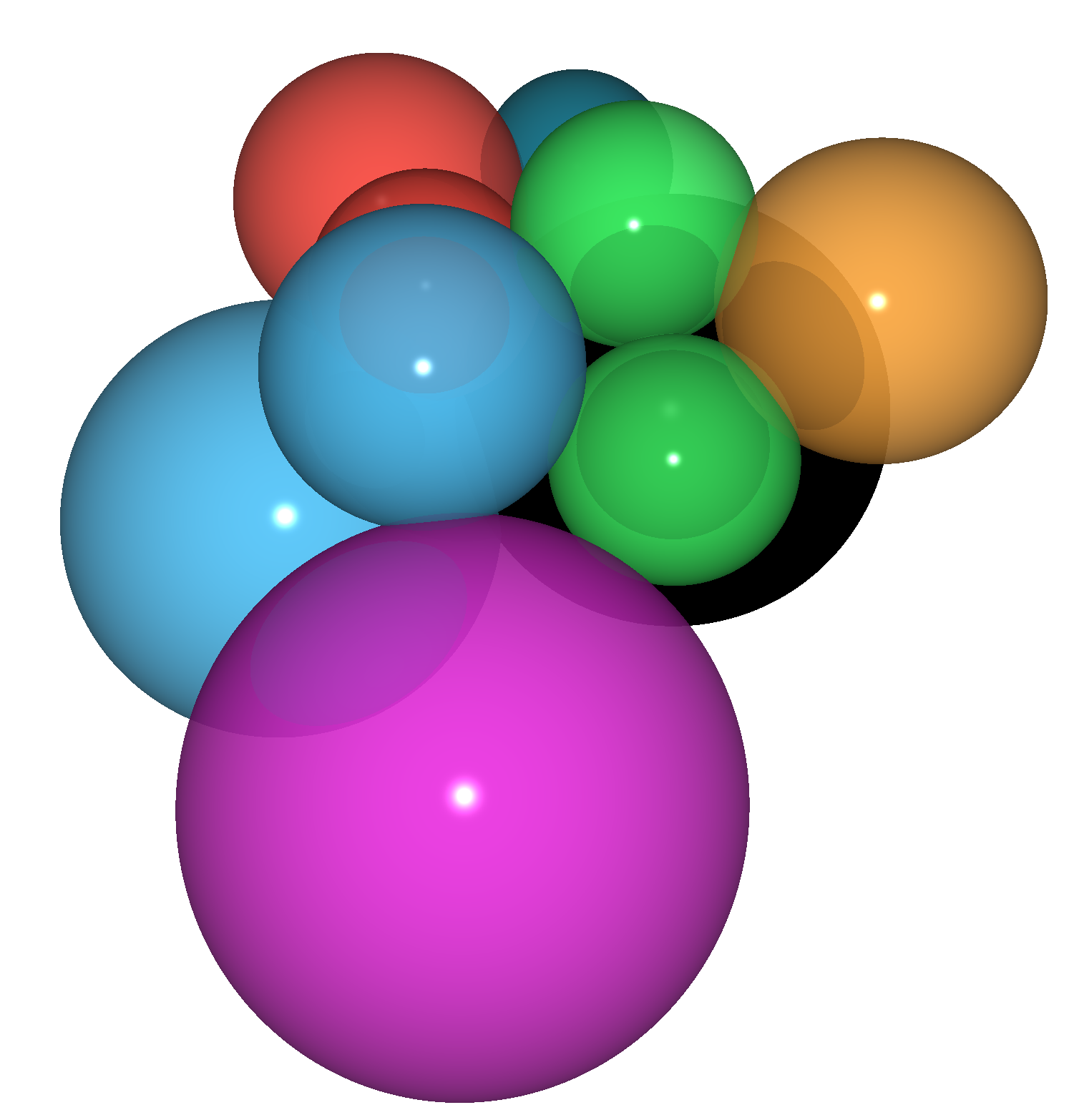}}
    \subfigure{\Large(b)}{\includegraphics[width=0.4\columnwidth]{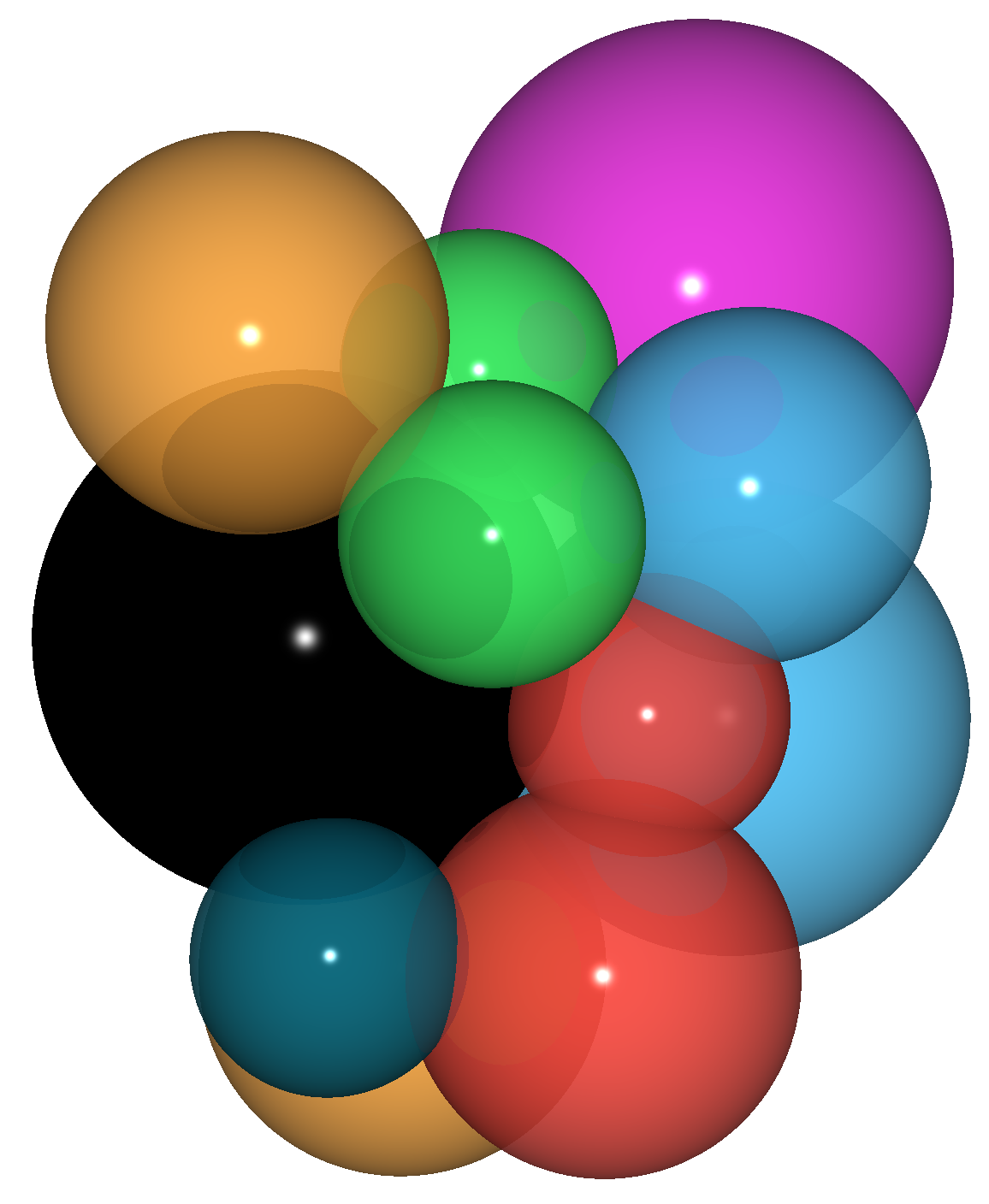}}
    \caption{Two views of the same central particle (black, $R=4.66$~$\mu${}m) that is part of 9 different tetrahedra and in contact with 10 particles. The particles are made slightly transparent for clarity. The large magenta particle has $R=4.98$~$\mu ${}m, in the front of (a) and back of (b). The smallest dark blue particle has $R=2.2$~$\mu${}m, barely visible in the top back in (a), and in the bottom front in (b).}
    \label{fig:cover_tet}
\end{figure}

In Fig.~\ref{fig:cover_tet}, we show an example with a central (black) particle which has ten particles in contact with it; the particle and its ten neighbors form nine tetrahedra.  This lends credibility to the physical idea of particles which land on the surface of the black particle and then can move until they contact other particles also on the surface, forming multiple contacts that are then more stable.  The neighboring particles of the central black particle have in a sense formed a two-dimensional aggregate on the surface of the black particle.

We wish to understand the geometric constraints that may be acting in colloidal gel formation with our highly polydisperse particles.  Any four \textit{monodisperse} spheres are guaranteed the \textit{possibility} of forming a tetrahedron. Unlike monodisperse particles, this guarantee no longer holds for polydisperse particles.  We consider the question: given four random spheres of radii $\{r_1,r_2,r_3,r_4\}$, what are the conditions on these radii such that they can form a tetrahedron?  A general tetrahedron is made of four triangular faces with six edges. Geometrically, for each face, the three edges must satisfy a triangle inequality. In addition, all four triangles must form a closed tetrahedron.  This latter condition can be tested by taking the volume of the tetrahedron via the Cayley-Menger determinant (a matrix whose elements are the edge lengths of the tetrahedron). If positive, the edge lengths can form a valid tetrahedron \cite{wirth_edge_2009}. The combination of the four triangle inequalities (triangle condition) and the Cayley-Menger determinant (closure condition) produces a ``tetrahedral inequality".  This inequality is, however, an inequality acting on six independent edge lengths.  For a tetrahedron made of spheres with radii $\{r_1,r_2,r_3,r_4\}$, the edges of the triangles are composed of the sum of the combinations of radii, meaning that a 4-sphere tetrahedron has 4 degrees of freedom rather than 6.

Considering the case of four spheres, we first consider the triangle inequality generated by three spheres.  The centers of any three spheres form a plane, so really this is the problem of three circles lying in a plane that mutually touch.  For all positive, real numbers any three combinations of circles produce a valid triangle.  That is, given three triangle edges $a=r_1 + r_2$, $b=r_2 + r_3$, and $c=r_1 + r_3$, the three triangle inequalities reduce to $r_1>0$, $r_2>0$ ,and $r_3>0$. Therefore, a 4-sphere tetrahedron always satisfies the triangle condition for each of its faces; thus the tetrahedral inequality reduces to the closure condition. Furthermore, by taking the determinant of the Cayley-Menger matrix with the six edge lengths generated from the four sphere radii, the closure condition reduces to a well known problem in mathematics known as the Interior Soddy Circle problem. That is, any three spheres can be positioned to touch; the important question is whether a fourth sphere can be added to that triangle such that it contacts all three spheres simultaneously. Thus, our initial tetrahedral inequality question reduces again, this time to a two-dimensional problem whose solution was first discovered by Rene Descartes and popularised by Frederick Soddy.  (The result was so beautiful that it drove Soddy to publish the poem ``The Kiss Precise" in \textit{Nature} \cite{soddy_kiss_1936}).  The result shows that the smallest the radius of the 4th particle can be before it no longer forms contacts is given by: 

\begin{equation}
    r_s=\frac{r_1r_2r_3}{r_1r_2+r_1r_3+r_2r_3+2\sqrt{r_1r_2r_3(r_1+r_2+r_3)}}.
    \label{eq:soddy}
\end{equation}
\noindent
$r_s$ is a ``Soddy radius" below which the addition of the fourth particle would fail to yield a tetrahedron:  that is, small spheres are the ones that will fail to touch larger spheres.  This is the lower bound on the small sphere radius; there is no upper bound, and any sphere with a radius larger than $r_s$ is able to touch the three other spheres.  If we consider the case $r_1 = r_2 = r_3$ to be three large particles, Eq.~\ref{eq:soddy} states that the range of particle sizes at which this restriction becomes relevant for determining possible tetrahedra is when the radius of the largest particles $3 + 2 \sqrt{3} \approx 6.7$ times the radius of the smallest particles. Prior work has studied the affinities between the Soddy circle problem and questions about 4-ball tetrahedral structures \cite{fox_soddy_2008}, but, to our knowledge, the route to the Soddy inequality through the Cayley-Menger determinant has not yet been reported.

For our particular sample, the tetrahedral inequality, reduced to Eq.~\ref{eq:soddy}, barely restricts the structure of the colloidal gel.  By randomly sampling our radius distribution, we find that the probability of crafting an invalid tetrahedron given four random radii is $P_{tet}^{theory}\approx 10^{-9}$; a value which is commensurate with the probability of choosing three particles of radii $> 8 \mu${}m.  That is, given that our size distribution (Fig.~\ref{fig:rad_dist}) has many small particles and few large particles, it is extremely rare that we would ever have three large particles mutually contacting each other, such that a small fourth particle would be unable to touch the three large particles.  We note that for a colloidal gel composed of a polydisperse sample where $P(R)$ has a more negative skew, such that the large particles are common and the small particles are rare, would be more subject to the Soddy radius constraint.  For example, if one draws four particles at random from a bidisperse distribution with sizes $R_a = 1, R_b = 7$ and probabilities $p_a = 1/4, p_b = 3/4$, then 42.2\% of the time the four particles would fail the Soddy radius inequality, and thus forming tetrahedra would be strongly influenced by this geometric constraint.

\begin{figure}
    \centering
    \subfigure(a){\includegraphics[width=0.9\columnwidth,viewport=119 399 532 687]{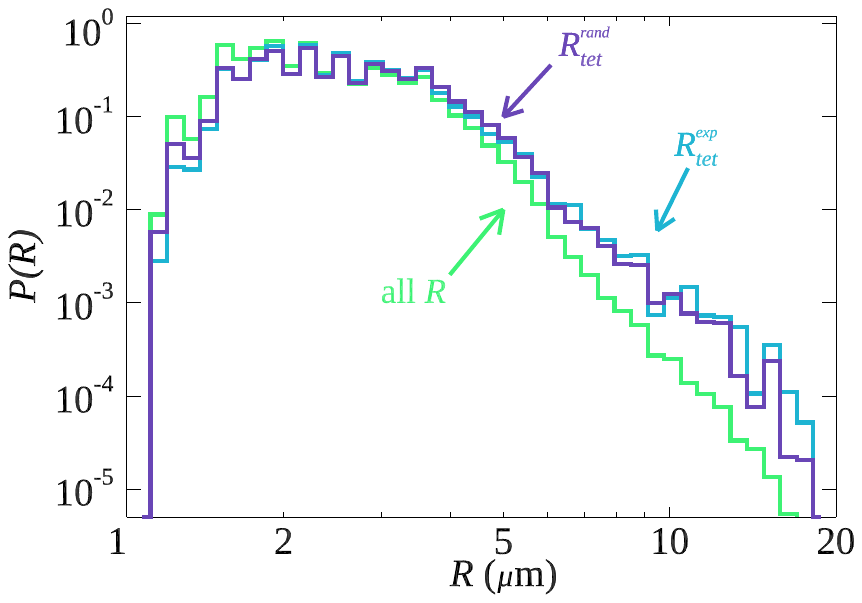}}
    \subfigure(b){\includegraphics[width=0.9\columnwidth,viewport=115 381 535 684]{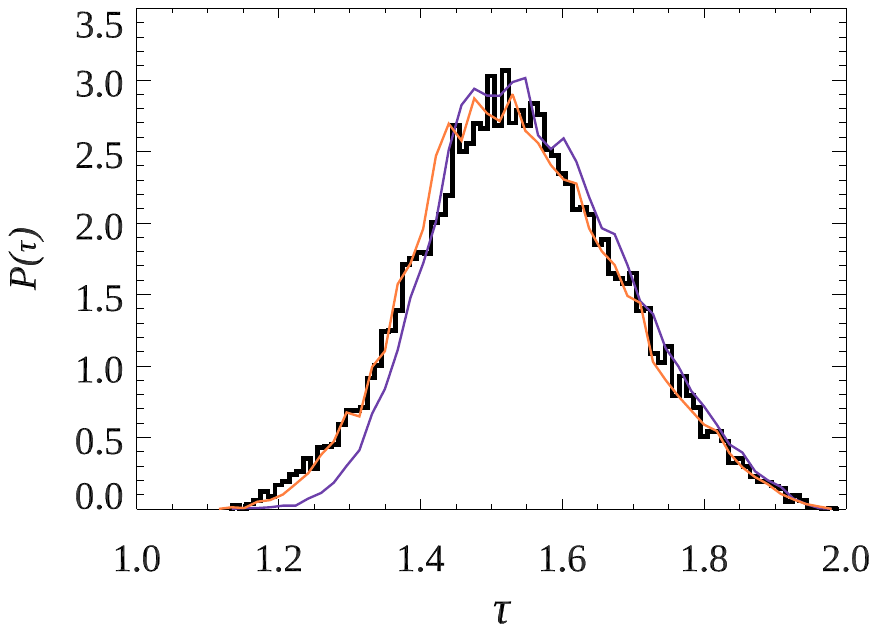}}
    \caption{ (a) Probability distributions of the radius distribution (green), the tetrahedron distribution (light blue), and the theoretical distribution of tetrehedra generated by randomly sampling the neighbor distribution and checking via Eq.~\ref{eq:soddy} for validity (purple). The contact and tetrahedron distributions are different from the radius distribution because they are allowed to contain repeated entries if a particle has multiple contacts or is a part of multiple tetrahedra. (b) Probability distribution of the asymmetry of tetrahedra, $\tau$. The rough black curve represents the experimentally measured tetrahedra, and the purple and orange curves represent tetrahedra formed by randomly sampling the radius distribution or the contact radius distribution respectively.}
    \label{fig:tet_calc}
\end{figure}

Despite the rareness of the geometric constraint applying in our sample, nonetheless the large particles play an interesting role in the formation of tetrahedra.  To characterize this, in Fig.~\ref{fig:tet_calc}(a), we plot the size distribution of all particles, particles that are neighbors, and particles that are in tetrahedra. The size distributions are different because particles can be repeated if they have multiple neighbors, or are parts of multiple tetrahedra.  The curve with the highest probability in the tail for large $R$ (blue) is the one corresponding to all particles found in tetrahedra.  Thus, while large particles are rare in our samples, Fig.~\ref{fig:tet_calc}(a) shows they have an out-sized presence in the formation of tetrahedra relative to their presence in the sample.  This is consistent with Fig.~\ref{fig:contact} which shows large particles have more neighbors in general, so a larger possibility of participating in tetrahedra.  In Fig.~\ref{fig:tet_calc}(a) the middle (violet) curve represents $P(R)$ for all contacting particles, which is quite similar to the $P(R)$ for contacting particles which are in tetrahedra, although nonetheless the tetrahedral particles have slightly more participation from the large particles.  Notably, both these curves are distinct from the bottom green curve which shows $P(R)$ for all particles.  This observation of the relative prevalence of large particles in tetrahedra may help understand Cantor \textit{et al.}'s computational observation that large particles help form a stress-resistant backbone \cite{cantor_rheology_2018}.

Finally, we consider the shapes of the tetrahedra formed in our highly polydisperse sample.  For each triangle in a tetrahedron, consider the edge lengths $a$, $b$, and $c$, with $c$ being the longest.  We define the parameter $\tau = (a+b)/c$.  Given that $c \geq a$ and $c \geq b$, and the triangle inequality requires $c > a + b$, then $\tau$ is bounded by $1<\tau\leq 2$.  $\tau=2$ corresponds to a completely monodisperse triangle, and increasing polydispersity within the triangle leads to $\tau \rightarrow 1$. Thus, each tetrahedron is composed of four $\tau$, which we then take the minimum of (to identify the least symmetric triangle) to assign each   particular tetrahedron an asymmetry parameter $\tau$.  The distribution of $\tau$ is shown in Fig.~\ref{fig:tet_calc}(b) as the noisy black curve.  The minimum at $\tau=2$ reflects the rarity of finding four particles of close to the same size to form a tetrahedra.  The thin solid violet curve is a null hypothesis for $P(\tau)$ constructed by randomly sampling four particles from $P(R)$ and forming a tetrahedron.  The experimental data clearly have more asymmetric tetrahedra than the null hypothesis curve.  The over-representation of large particles in Fig.~\ref{fig:tet_calc}(a) subsequently produces more asymmetric tetrahedra as shown by the low $\tau$ tail in Fig.~\ref{fig:tet_calc}(b). We can correct for this over-representation by generating tetrehedra from the contact radius list, which includes repeats if particles form multiple contacts. In Fig.~\ref{fig:tet_calc}(b), we show this result in the thin light orange curve which fits the experimental curve well at low $\tau$.

\section{Conclusion}

We have studied the structure of colloidal gels composed of particles that span more than a decade in radii.  Larger particles have more particles attached to them as compared to smaller particles.  This finding is reasonable given that the larger particles have a larger surface area; although we observe that the number of neighbors scales as $R^0.7$ and not $R^2$.  All potential particle pairs between particles of radii $(R_1,R_2)$ occur with a probability essentially proportional to $P(R_1) P(R_2)$ in terms of the sample's particle size distribution $P(R)$.  Given that the large particles are rare, this means that it is rare to see two large particles stuck together, but not impossible.  This could be controlled by changing the underlying particle size distribution to favor larger particles.

When two particles stick together, the contact is sufficiently ``loose'' that they have some ability to rearrange.  This results in the formation of tetrahedral structures within the colloidal gel, where four particles are mutually touching, resulting in a much more stable configuration that is much harder to rearrange \cite{royall_direct_2008,zia_micro-mechanical_2014}.  Given that large particles have more nearest neighbors, a consequence is that large particles participate in tetrahedra more than might otherwise be expected, and thus large particles frequently serve as rigid points of the gel.

Our observations are for our specific particle size distribution shown in Fig.~\ref{fig:rad_dist} where large particles are rare.  In our samples, large particles are more frequently associated with tetrahedral structures, and this result should be general.  In any colloidal gel, large particles will have the ability to have more nearest neighbors than smaller particles, and thus a large particle will have an increased chance that some of those nearest neighbors form a tetrahedron with the large particle.  However, we note that for a particle size distribution where the largest particles occur more frequently, we will have the Soddy radius restriction as a geometric constraint preventing some sets of four particles from being able to form a tetrahedron.  Those colloidal gels will, in some sense, be ``less random'' due to this geometric constraint.  This geometric constraint arises when the size ratio between largest and smallest particles is at least $\sim 6.7$.

Another potential difference between our highly polydisperse gels and other colloidal gel systems relates to the rearrangements we believe occur in our gels that create tetrahedral structure.  In a colloidal gel with stronger attractive energies (such that $k_BT$ is insufficient to break a bond) and an inability for particles to roll or slide against each other, it seems likely the average contact number would be smaller and tetrahedral structures scarce.  We would also expect that preventing internal rearrangements within the gel would decrease the fractal dimension from our observed value of $2.5 \pm 0.1$ \cite{liu_fractal_1990}.  For example, a colloidal gel composed of faceted particles that could not roll against each other found the fractal dimension was 1.8 to 1.9 depending on conditions \cite{rice12}.

A final difference to note is that many prior experimental studies used the depletion force to form the colloidal gel \cite{segre01prl,dinsmore_direct_2002,dinsmore06,dibble06,lee08,lu08,royall_direct_2008,rice12,pandey13,koumakis15,razali17,whitaker19,datta2025}.  The depletion force arises by adding small polymers to the colloidal sample \cite{asakura54,lekkerkerker92}:  the exclusion of polymers from between particles results in an unbalanced osmotic pressure, pushing the particles together \cite{evans98}.  The depletion force is primarily controlled by the polymer concentration, but additionally it depends on the particle size.  The larger the particles are, the larger the force \cite{asakura54}.  For example, the attraction energy of a small particle to a much larger particle is twice that of two small particles sticking together \cite{kaplan94,dinsmore96}.  Intriguingly, this same factor of two is true for any short-ranged interaction potential, as shown by the Derjaguin approximation \cite{israelachvilibook}; it is true for our experimental attractive van der Waals force. 
The size dependence of our attractive force (Eqn.~\ref{vdw}) could be biasing the formation of our colloidal gels.  On the other hand, given that the energy of attraction is much larger than $k_BT$, the size dependence of the van der Waals strength may be irrelevant, in contrast to depletion forces which are more typically $O(k_BT)$.

\begin{figure}[b]
    \centering
    \includegraphics[width=\columnwidth]{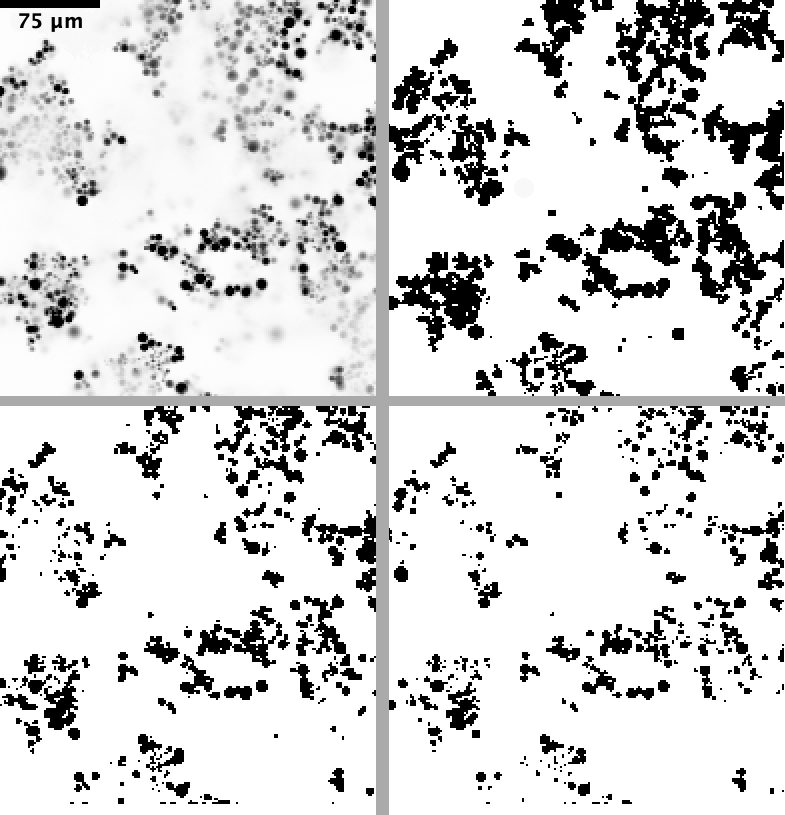}
    \caption{Three different global thresholds applied to the same image in the top left. The threshold increases from top right to bottom right. The image in the top left is a $\sim 300 \times 300$~$\mu$m$^2$ 2-dimensional slice of a larger $712\times712\times105$~$\mu$m$^3$ volume used for the determination of $D_f$. From top right to bottom right, $D_f$ ranges from $D_f=2.61$ to 2.45.  The bottom left image is the threshold level used for our particle detection.}.
    \label{fig:thresh_frac}
\end{figure}

\section*{Acknowledgments}
We thank A.~B.~Schofield for synthesizing our colloidal particles and thank J.~C.~Crocker, W.~C.~K.~Poon, A.~B.~Schofield, and M.~Wilson for helpful conversations.  Our work was supported by NASA (Grant No.~80NSSC22K0292) and in part by the Emory University Integrated Cellular Imaging Core of the Winship Cancer Institute of Emory University and NIH/NCI under award number, 2P30CA138292-04 (RRID:SCR\_023534). The content is solely the responsibility of the authors and does not necessarily reflect the official views of the National Institute of Health.

\section*{Data Availability}

The data that support the findings of this article are openly
available \cite{dataonline}.

\section{\label{sec:Appendix} Appendix}

An artifact of using the box-counting fractal dimension algorithm on real space fluorescence microscopy images is the smooth, monotonic dependence of $D_f$ on the chosen threshold. Our initial estimate for a reasonable threshold is given by the same threshold used for particle detection. Our error is determined by tuning the threshold above and below this value until we see either the erosion of in-focus colloids or the erasure of network detail typical of under-thresholding. We show a representative example of this process in Fig.~\ref{fig:thresh_frac}.

\end{document}